# BIMxAR: BIM-Empowered Augmented Reality for Learning Architectural Representations

Ziad Ashour
Department of Architecture
Texas A&M University
College Station, TX 77840, USA
zashour@tamu.edu

Zohreh Shaghaghian
Department of Architecture
Texas A&M University
College Station, TX 77840, USA
zohreh-sh@tamu.edu

Wei Yan
Department of Architecture
Texas A&M University
College Station, TX 77840, USA
wyan@tamu.edu

**Abstract**

Literature review shows limited research investigating the utilization of Augmented Reality (AR) to improve learning and understanding architectural representations, specifically section views. In this study, we present an AR system prototype (BIMxAR), its new and accurate building-scale registration method, and its novel visualization features that facilitate the comprehension of building construction systems, materials configuration, and 3D section views of complex structures through the integration of AR, Building Information Modeling (BIM), and physical buildings. A pilot user study found improvements after students studied building section views in a physical building with AR, though not statistically significant, in terms of scores of the Santa Barbara Solids Test (SBST) and the Architectural Representations Test (ART). When incorporating time as a performance factor, the ART timed scores show a significant improvement in the posttest session. BIMxAR has the potential to enhance the students' spatial abilities, particularly in understanding buildings and complex section views.

1. **Introduction**

Augmented Reality (AR) is being researched in the education sector [1], [2] about its tangible benefits including increased spatial abilities, learning gains, motivation, and collaboration [3], [4]. In architecture education, AR has been used to train students to produce orthographic projections and understand building components [5]–[7].

The ability to translate virtual information and relate it to the physical world is a crucial skill in the domain of architecture. Cognitive mental loads on the students are anticipated during the process of translating and relating components of a 2D or 3D drawing to their locations in the physical world due to the differences in views, perspective angles, and scales [8]. The mental effort required to process multiple sources of information that are distant from each other can increase the extraneous cognitive load [9]. One key feature of AR is superimposing virtual content relative to its correct location in the physical world.

This research seeks to explore the AR effects on assisting students to comprehend and reproduce architectural sections by utilizing AR – augmenting physical buildings by virtual building models. There is scant research investigating the utilization of AR in facilitating learning and the creation of building sections, which are important in building design, construction, and modeling. Additionally, the limited examples of BIM-enabled AR in the literature lack the level of interaction needed for building components inspection. Thus, further investigation in these particular areas is required. The research asserts the necessity to explore new methods that improve spatial abilities in the domain of architecture education. Moreover, the research is expected to contribute to architectural education and the body of knowledge by suggesting a learning approach for students to comprehend building construction systems, and materials assembly and configuration. The study intends to support the students' understanding of section views of complex structures. The current research showcases the performance and the technical aspects of our working prototype (BIMxAR) towards this approach. We present: (1) the workflows, (2) the model registration methods that we have explored and developed in BIMxAR, (3) extraction of the Building Information Modeling (BIM) metadata and its utilization in AR, (4) the user interface and the graphical representations inside an AR environment, (5) user interaction with the AR environment, and (6) the section creation function. Furthermore, the current study presents the results of a pilot user study that was conducted to measure the participants' learning gain and their mental cognitive load while using the prototype.

## 2. Related Work

### 2.1 Augmented Reality (AR)

Virtuality continuum (VC) is a continuous scale spanning from a real-world to a virtual environment, and anything in between is a potential combination of real and virtual objects (mixed realities). One such combination is AR [10]. Unlike Virtual Reality (VR), where the user is completely immersed in a synthesized environment, which is disconnected from the real world around the user, AR enhances real-world perception by complementing it with virtual objects [11]. AR can be defined as an interactive display system that enhances reality in real-time by contextually aligning virtual objects with the physical world [12]. Physical environment tracking and virtual information registration in the real world are the key functions of an AR system [13]. The tracking unit in an AR system must understand the environment and track the camera relative to the real world in order to correctly align virtual information with a real-world environment [14]. The selection for a tracking or registration method depends on the application it will be used for and the environment it will be used in [15]. Registration methods can be categorized into three categories: vision-based, sensor-based, and hybrid methods [16].

### 2.2 Spatial Ability Training with AR

Spatial ability is the human ability to mentally manipulate an object to represent it through a different viewpoint [17]. Spatial ability is strongly correlated to the academic performance of students, particularly students studying STEM (Science, Technology, Engineering, and Math) subjects [18]–[22] and AEC [23]. Spatial ability includes spatial visualization, which involves multiple complex tasks; spatial relations, which involves simpler tasks, e.g., fast mental rotations; and spatial orientation, which involves representing an object from a different perspective [24]. Spatial visualization and orientation are important skills for architecture and construction students. Moreover, they enable students to remember the built environment's organization and structure [23].

Studies have shown that students' spatial abilities can be improved with special training [25], [26]. Various studies demonstrated the positive impact of AR and VR in improving students' spatial abilities [17], [21], [24], [27]–[30]. A study explored spatial memory development and how spatial knowledge is acquired through the use of VR [31]. Shi et al. [31] investigated spatial memory development and how spatial knowledge is acquired through the use of VR and their findings asserted the role of visual context (3D and VR) for developing spatial memory and the strong correlation between spatial memory and visual attention. Moreover, Sanandaji et al. [32] investigated VR utilization in spatial training to better understand 2D cross-sections of complex structures. The study documented improvements in abilities, such as cross-sections, mental rotations, and viewpoint visualization. Dünser et al. [17] explored the difference between AR and VR in improving spatial ability and concluded that AR could be advantageous in certain tasks that include objective perspectives. In their study, although AR showed improvement, yet minor, in mental cutting tasks, no significant difference was observed when compared to a non-AR training tool (computer screen) using the Mental Cutting Test (MCT) [33]. Moreover, while their study recruited 215 participants and made them undergo lengthy repetitive training sessions, they concluded that AR did not provide any clear evidence of spatial ability improvements. However, Ali et al. [34] found that the experimental group who underwent spatial ability training using AR was significantly better than the control group in mental cutting and folding abilities. Furthermore, previous research by Contero et al. [30] showed that the group who received spatial ability training using AR performed significantly better than the control group in the Mental Rotation Test and Differential Aptitude Test (Spatial Relations). Additionally, many studies, such as [35] and [36], have shown how AR can reduce the completion time of tasks that require spatial abilities. Due to the inconsistent results about AR's impacts on spatial training, more research in this field is needed.

### 2.3 Extraneous Cognitive load

Extraneous cognitive load is the mental effort exerted by the learner to process the presentation and design of instructional materials. Increased extraneous cognitive load can negatively affect learning and increase the overall cognitive load [37]. Additionally, the mental effort required to process multiple sources of information that are distant from each other can increase extraneous cognitive load [9]. The AR capability of superimposing virtual information on its relative location in the physical world can reduce extraneous cognitive load and ultimately enhance the learning process [38].

Although AR has great potential and benefits in education, instructional materials have to be effective and well-presented within the AR environment to avoid increased cognitive load due to the learning content complexity [1], [39].

### 2.4 AR in Architectural Education

AR has the potential to reform the architecture, construction, and engineering education [40]. It has been already explored in several areas in architecture and construction education. For example, AR has been employed in project presentation [2], [41], design [42], teaching CAD [5]–[7], [43], geometric transformations [44], [45], architectural history [46], structural analysis [47] and architectural lighting [48]. In spite of that, our review of the literature indicates a little emphasis on the utilization of AR in teaching students building construction system integration, material assemblies, and section view creation. Moreover, many studies, such as [40], [49], [50], lack the alignment of the virtual and the physical building objects, which is a core feature of a true AR experience. Little examples in the literature utilize this core feature. Additionally, the amount of interaction that allows students to inspect the virtual content (building components) is very limited, in examples such as [49], [51]. Furthermore, other examples, such as [52], provide limited visualizations through axonometric views from one single angle.

### 2.5 Tests, Test Scores, and Completion Time

For evaluating the impacts of the developed BIMxAR system on spatial and architectural learning, we have conducted a pilot user study that examining students' learning gains reflected by the Santa Barbara Solids Test (SBST) and our designed Architectural Representations Test (ART), considering both test scores and completion time. When incorporating time with score as a performance factor, it could provide us a more detailed understanding of the student's performance and abilities, and construct a profile to show his or her strengths and weaknesses [53]–[56]. However, utilizing time and score could result in complicated implications when drawing conclusions. Faster or slower responses could be interpreted incorrectly. For example, faster responses might occur because some students utilize guessing as a strategy to answer questions, or simply lost motivation in the test subject [57]. Moreover, slower responses could be explained as students being careful, having a slow pace in solving questions [57], or suffering from language difficulties [54]. Different analysis models have been proposed to handle completion times and scores, such as item response theory and cognitive diagnostic models [58]. Some of these models could be used to detect and solve some of the related tradeoffs, such as cheating, time management habits or behaviors, motivation levels, and solving strategies [57]. Additionally, more control measurements were suggested to counter the issues related to using time and scores, including data collection of eye movements, EEG and brain imaging, and number of clicks or moves on the computer's screen [55]. In our project, performance evaluations using test scores and completion time are conducted to provide a more comprehensive assessment for the AR-assisted learning outcomes.

## 3. Methodology

The present research seeks to build and test an educational tool that supports architectural students' comprehension of building construction systems, material assemblies and configurations, and architectural representations. The developed prototype utilizes the physical-virtual overlay feature to facilitate spatial learning using existing physical buildings and their Building Information Models (BIM). The design of the AR prototype takes into consideration the different benefits that can be provided by this overlay feature. The design makes use of this feature to superimpose BIM information (geometrical and textual information) on the physical built environment. Additionally, it enables the user to virtually cut and see through the building being inspected, in a way similar to magnetic resonance imaging (MRI), to provide better visualization that enables embodied learning for improved understanding of the internal elements behind finishes and how they integrate with other systems. The study assesses AR's effects on improving architectural education through a pilot user study.

The current research consists of two parts. The first part showcases the performance and the technical aspects of BIMxAR in terms of the workflows, registration methods, BIM metadata extraction and retrieval in AR, user interface and graphical representation, and section view creation. The second part presents the pilot user study that was conducted to: (1) measure participants' learning gain in subjects including the mental cutting abilities and the understanding of the architectural representations, and (2) measure the participant's mental cognitive load while using the prototype, using a subjective survey.

### 3.1 *BIMxAR Development*

To develop BIMxAR, we used Unity [59], which is a common AR platform and gaming engine. Unity houses AR Foundation, which contains core features of ARCore and ARKit. Programming in Unity was done using the C# language, and the developed prototype is an iOS application deployed to devices including iPhone 7 Plus, iPhone 12 Pro, and iPad (8$^{th}$ Generation).

### 3.1.1 *AR Model Registration*

We considered three different solutions to register the virtual model in the physical space. The first solution is based on our previous prototype, which utilizes GPS, inertial measurement unit (IMU) sensors, and manual transformation controls [60]. The solution uses GPS and IMU sensors during the initial registration outside the building and the transformation sliders in the user interface that can be used to manually correct the registration before entering the building. Once the model is correctly registered, the application will only depend on the IMU sensors. The registration method was tested using the BIM (Revit) project and the physical building of the Memorial Student Center (MSC) building on the Texas A&M University campus in College Station, Texas. The prototype was an iOS AR application deployed to the iPhone 7 Plus.

The performance of the first solution was tested to assess whether it can maintain the virtual model alignment with the physical building. Since the device only depends on the device's IMU when indoor, minor misalignment (drift) of the virtual model was expected due to the accumulated errors in the IMU calculations. Four tests were conducted to verify the alignment performance and measure the drifting errors; all of the tests were carried out after the virtual model was correctly aligned with the physical building. The average drift was approximately between +/- 0.9 m to +/- 2.0 m. The solution can support registration in outdoor and part of indoor environments, but accurate alignment cannot be maintained in the indoor environment when relying only on the device's motion sensors, as shown in **Figure 1**. This necessitates the integration of other types of tracking, such as computer vision and Artificial Intelligence, specifically deep learning methods.

The second solution utilizes computer vision and the point clouds of the pre-scanned physical space as a registration method. It was tested using two different scanners: (1) the iPhone 12 Pro Max built-in light detection and ranging (LiDAR) scanner, and (2) the Matterport Pro2 scanner. We employed Vuforia Area Target [61] to accomplish the registration.

The method using the first scanner requires scanning the physical space through the Vuforia Area Target Generator App and then processing the collected data remotely in the Vuforia Cloud. Once the data are processed, a database will be generated and then imported in Unity to be used by the Vuforia Engine.

The method using the second scanner requires scanning the physical space at different scanning points through the Matterport Capture App. Then, uploading the scan file to the Matterport cloud to be processed and to generate the MatterPak package that will be utilized in the Vuforia Area Target Generator App to generate the database. Once the database has been generated, it is imported in Unity to be used by the Vuforia Engine.

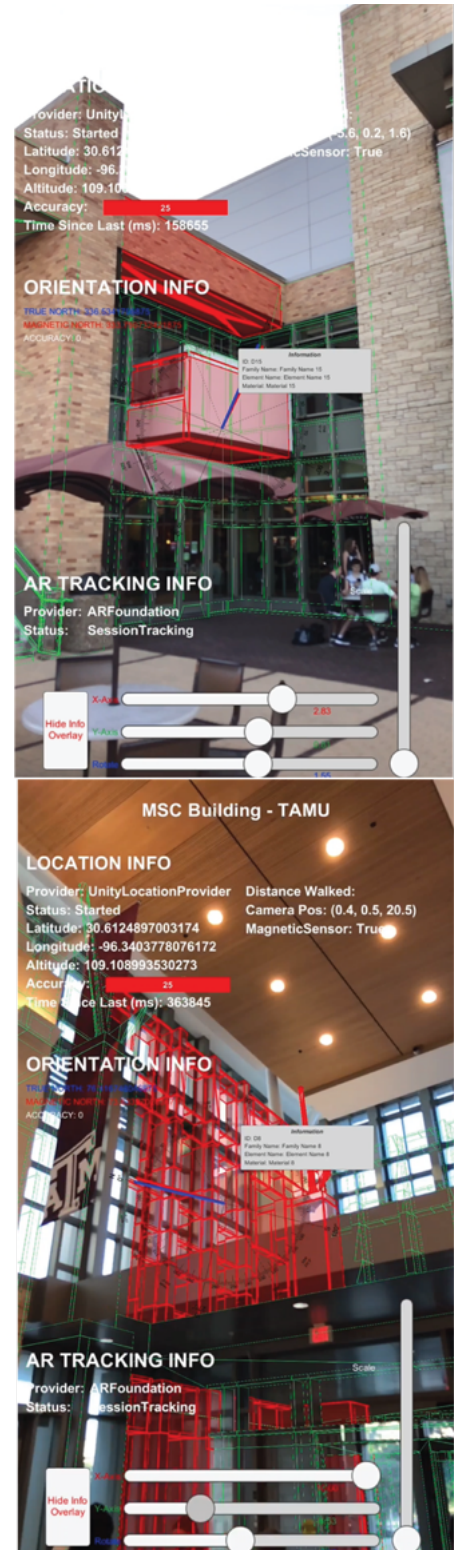

**Figure 1:** Testing the first registration method at the MSC building using GPS and IMU sensors. Red and green wireframes show the BIM models. Top: Registration outside the building; Bottom: Registration inside the building.

Although both scanning methods provide fairly accurate and robust registration, the scan of the physical space is not sufficiently accurate at corners and edges (rounded instead of sharp), making it difficult to accurately align the virtual model with the scanned space in Unity as shown in **Figure 2** and **Figure 3**. Moreover, the rounded corners of the walls in the scanned space made BIMxAR suffer from misalignment issues whenever the user approach a corner. Our experiments suggested that the misalignment was due to the reason that Vuforia Area Target was constantly trying to match the corners and edges seen by the AR camera with the scanned space.

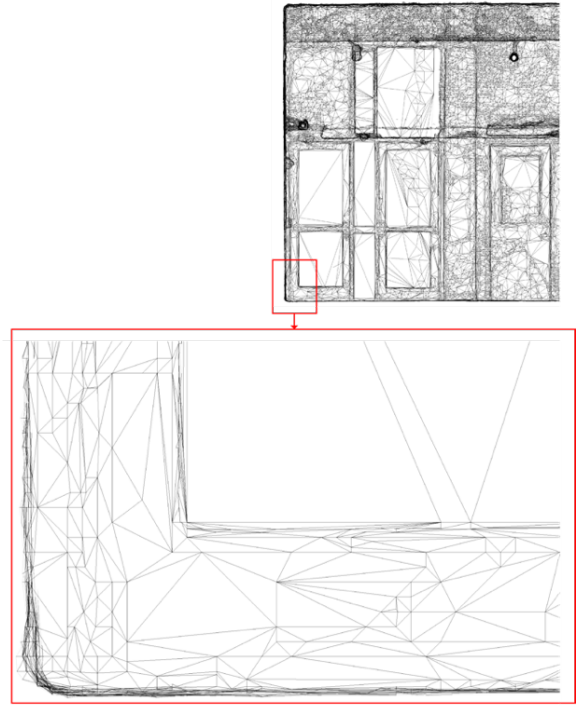

**Figure 3:** Scan quality using the Matterport Pro2 scanner. Corners of the walls are still rounded instead of sharp.

The third solution utilizes computer vision and 3D model-based AI/Deep Learning (DL), e.g., Vuforia Model Target. We employed Vuforia Model Target, which is normally used for registering small-scale 3D objects, e.g., artifacts and cars, but not designed for registering large environments, such as a space or building, in the physical environment.

The adopted method requires an accurate reconstructed 3D model of the physical building in order to generate a model target database in Vuforia Model Target Generator (MTG) that will later be utilized by BIMxAR to recognize and track the physical building, as shown in **Figure 4** (**Right**) and **Figure 5** (**Bottom**). Through extensive experiments, we adopted 3D model-based Deep Learning with 3D Scanning-corrected BIM (DL-3S-BIM) as our registration method for the scale of buildings, and this method has been proven to provide the best solution in terms of accuracy and robustness, as shown in **Figure 5** (**Bottom**).

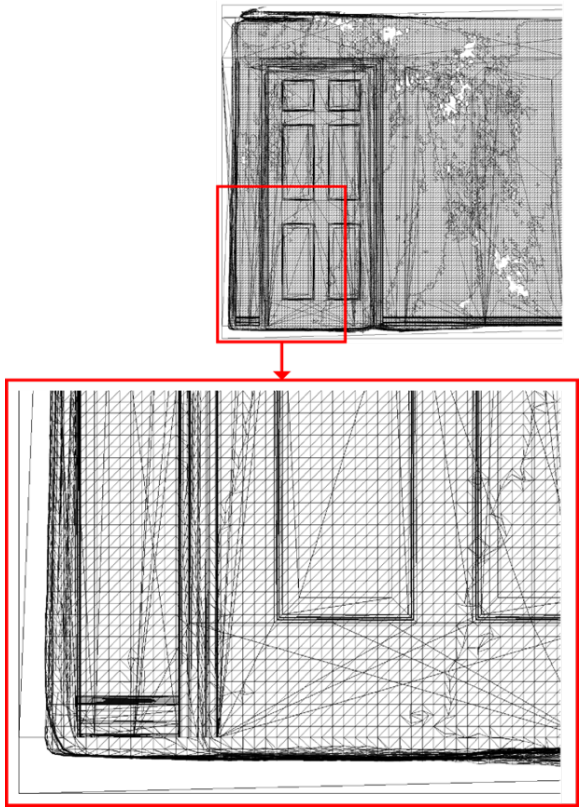

**Figure 2:** Scan quality using the iPhone 12 Pro Max built-in LiDAR sensor. Corners of the walls are rounded instead of sharp.

The reconstructed 3D model in **Figure 4** (**Middle**) was created based on measurements taken manually. We noticed that the generated model target database from the 3D model based on manual measurements could only work with small spaces. However, in larger spaces, the 3D model of the physical space must be based on more accurate measurements using a professional 3D-Scanner, e.g., Matterport's Structured Light Scanner. Therefore, we used the Matterport Pro2 scanner to scan the entire space and reconstruct its 3D model, as shown in **Figure 5** (**Middle**).

During the training, a cone view (virtual camera view) in Vuforia MTG is placed inside the 3D model. The location is defined to cover most of the physical space by setting the cone view at the midpoint of the space height. For smaller spaces, the azimuth range was set at 360 (degrees), and the elevation range from -40 to +50 (degrees), as shown in **Figure 6**. For larger spaces (DL-3S-BIM), the azimuth range was set at 360 (degrees), and the elevation range from -90 to +90 (degrees), as shown in **Figure 7**. The last step is to align the 3D model with the generated database (target model) in Unity to enable BIMxAR to spatially register the 3D model in its correct location and orientation in the physical world.

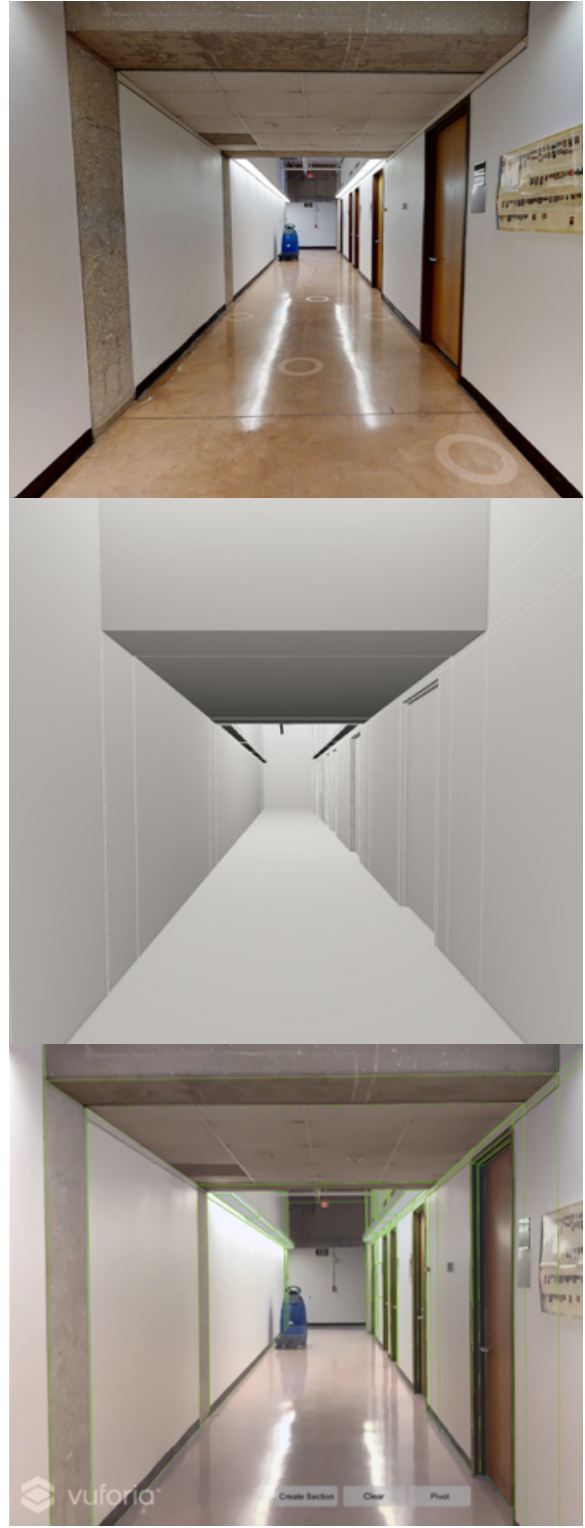

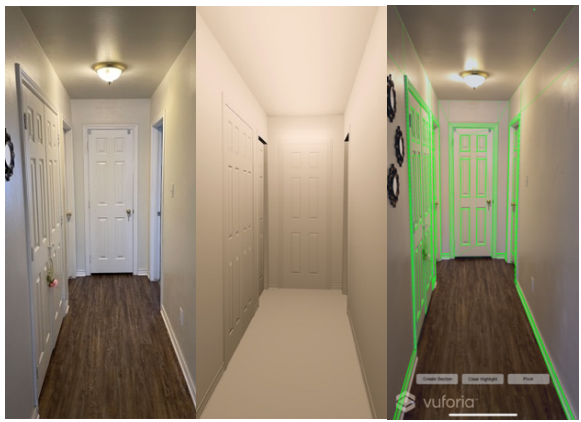

**Figure 4:** Left: actual physical space, middle: 3D BIM model (shaded) of the physical space based on manual measurements (built for Vuforia DL), right: the BIM model (wireframe) registered in the physical space.

**Figure 5:** Top: actual physical space; Middle: 3D BIM model (shaded) created manually and corrected with Matterport Pro2 3D scan of the physical space (built for Vuforia DL); Bottom: the BIM model (wireframe) registered in the physical space with high accuracy using the DL-3S-BIM method.

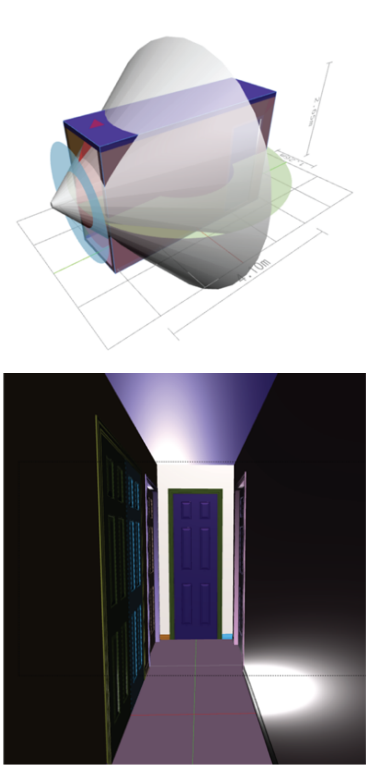

**Figure 6:** The imported virtual model, based on manual measurement of the physical space (in FBX format) in Vuforia MGT. The Azimuth Range is set to 360 degrees, and the Elevation Range is set from -40 to +50 degrees.

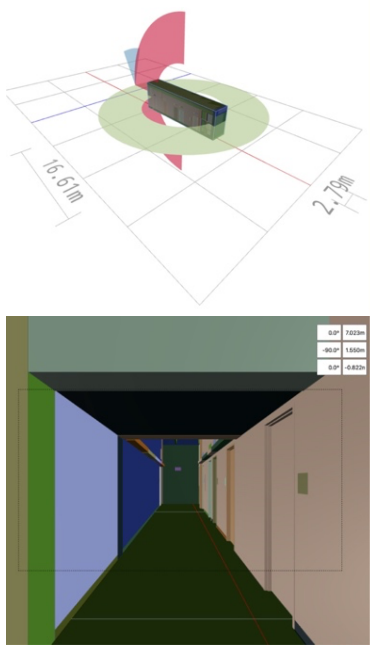

**Figure 7:** The imported virtual model, corrected with Matterport Pro2 3D scan of the physical space (in FBX format) in Vuforia MGT. The Azimuth Range is set to 360 degrees, and the Elevation Range is set from -90 to +90 degrees.

The Vuforia Model Target method for smaller spaces (using a virtual model based on manual measurement) and the DL-3S-BIM method were evaluated to quantitatively measure BIMxAR performance in registering the virtual model in the physical environment. The error of the registration is defined as the distance measured in the 2D projection of the 3D edges of the physical building and its virtual model. The error is not defined and measured as the 3D distances of the edges, because the measurements are 2D view-dependent and not truly measurable in 3D. The analysis was conducted by visually inspecting multiple screenshots (2D perspective images) and manually measuring the distances at the edges between the virtual model and the physical building. The manual measurement was done in Adobe Illustrator by first creating a vertical line representing the real height of the physical building, then scaling the screenshot image to match the corner-edges of the physical building with the vertical line, and finally measuring the difference between the virtual model and the physical space, as shown in **Figure 8** and **Figure 9**. The average error throughout the virtual model in smaller spaces (manual measurement) is around 15.7 mm and 15.00 mm when using the DL-3S-BIM.

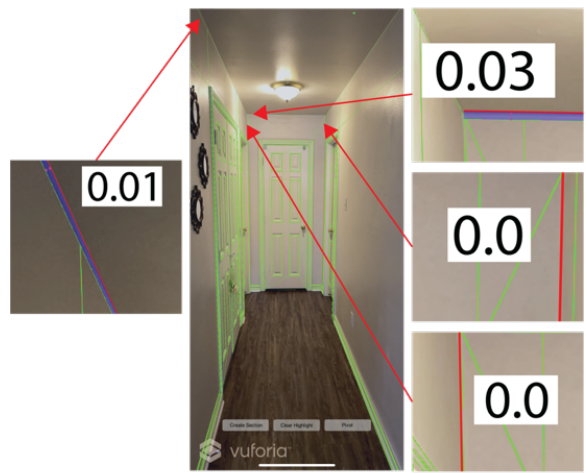

**Figure 8:** Measuring the performance of the registration method (Vuforia Model Target) in Adobe Illustrator. The figure shows the alignment differences (in meters) at multiple locations in one of the screenshots.

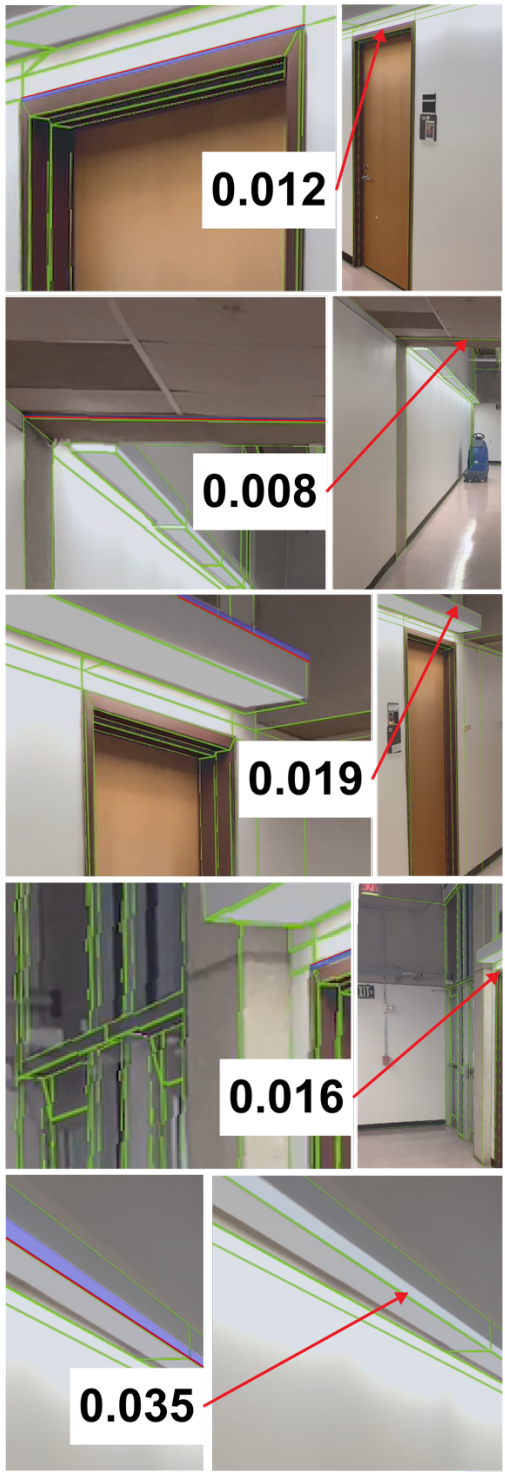

**Figure 9:** Measuring the performance of the registration method DL-3S-BIM in Adobe Illustrator. The figure shows the alignment differences (in meters) at multiple locations in one of the screenshots.

*3.1.2 Workflow*

The workflow utilizes BIM (Revit) files, in which the geometric and non-geometric information can be both accessed in Unity, as seen **Figure 10**. The geometric information (3D model) is exported as an FBX file format (while preserving the building components' IDs) to be used in Vuforia MTG and Unity. The extraction of BIM metadata is accomplished through Dynamo (a visual programming tool for Revit), as seen in **Figure 11.** The proposed approach collects the building model metadata, including the building components' IDs, categories, families, and all related parameters, and exports them into a CSV file format. The CSV file is then converted to the JSON format in order to be stored in a real-time database (Firebase). A script was developed to enable Unity to retrieve building objects' metadata through their IDs directly from the real-time database.

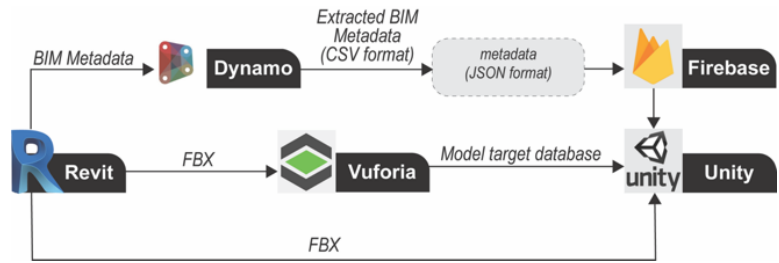

**Figure 10:** Workflow BIM (Revit) to AR development in Unity

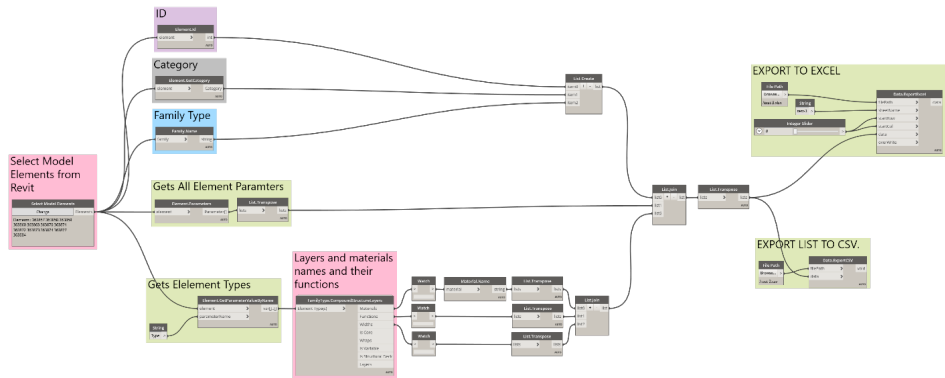

**Figure 11:** Approach for extracting BIM metadata from Revit via Dynamo.

### 3.1.3  User Interaction

***Materials:***

The first consideration in user interaction with the virtual world in the AR setting is the selection of the virtual model shaders and materials. By default, the assignment of opaque shaders for the virtual model will always occlude the physical building on the AR screen, no matter what spatial relations (front or back) exist between the virtual and physical objects.

To handle this AR occlusion problem, we decided to use a transparent yet occlusive shader highlighted with a wireframe and assigned it to the virtual model as used in [62]. As a result, the user can simultaneously view the physical and virtual objects with correct occlusions between them - objects in front occlude those on the back, no matter the objects are physical or virtual.

***BIM Metadata Retrieval:***

The second consideration is to enable the user to retrieve information about a building component or element. If the user touches an object of interest, it will be highlighted with a red wireframe shader, and a table of relevant information will be displayed, as shown in **Figure 12**.

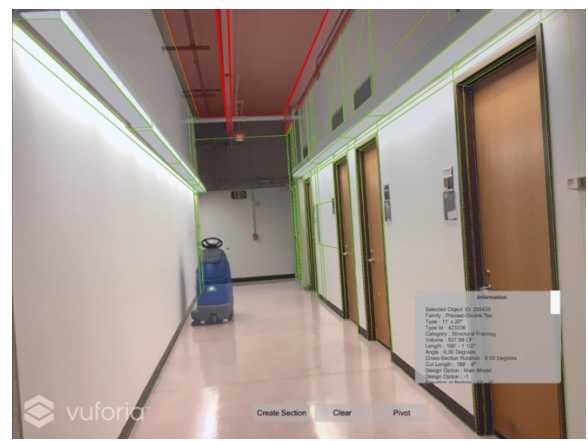

**Figure 12:** BIM model (green) and highlighted BIM object (red) with related BIM metadata table.

***Section Mode:***

The section creation function allows the user to spatially slice the building to create architectural section views. When creating a section, BIMxAR does not change the geometry. Instead, a Unity Asset shader, named crossSection [63], is adopted to create a rendering effect that can be designed to show the section views. The shader allows BIMxAR to create sections by only rendering the part behind the sectional plane and the rest of the model in front of the plane is hidden. It also provides hatch patterns for the section poche. The previous examples in the literature review enable a user to examine a building from specific section views, but preventing the user from examining other parts of the building or revealing internal building elements at specific locations, and thus the user cannot fully inspect the internal parts. In contrast, BIMxAR enables the user to freely control the sectional plane location and orientation, allowing the user to inspect the building from different architectural section views, supported by other advanced visualization features described in User Interaction.

BIMxAR contains six sectional planes to create a bounding box that surrounds the virtual model or a part of it. This configuration enables the user to create sections at all three axes (X, Y, and Z) with two orientations (left-right / front-back). To control the location of the sectional planes, the interface has three pairs of translation sliders (X, Y, and Z). **Table 1** shows the six translation sliders of the section plane and their functions. Also, multiple (up to three) sectional views can be simultaneously viewed to inspect the model from different sides.

**Table 1:** Translation sliders of the sectional plane

| Slider | Function |
|---|---|
| X-Axis (Pos) | Translates the sectional plane towards the positive direction of the *X-Axis* and the sectional plane normal is facing the *negative* direction. |
| X-Axis (Neg) | Translates the sectional plane towards the negative direction of the *X-Axis* and the sectional plane normal is facing the *positive* direction. |
| Y-Axis (Pos) | Translates the sectional plane towards the positive direction of the *Y-Axis* and the sectional plane normal is facing the *negative* direction. |
| Y-Axis (Neg) | Translates the sectional plane towards the negative direction of the *Y-Axis* and the sectional plane normal is facing the *positive* direction. |
| Z-Axis (Pos) | Translates the sectional plane towards the positive direction of the *Z-Axis* and the sectional plane normal is facing the *negative* direction. |
| Z-Axis (Neg) | Translates the sectional plane towards the negative direction of the *Z-Axis* and the sectional plane normal is facing the *positive* direction. |

Multiple tests have been conducted to examine the visualization performance of BIMxAR in an AR environment. During the section creation mode, we noticed if a large portion of the model is discarded, or more than one section view is created, the user cannot know if the virtual model is still correctly registered in the physical environment. Therefore, we decided to include the discarded part of the model in the rendering pipeline during the section creation mode. The discarded part is rendered with a completely transparent shader highlighted with a wireframe, as shown in **Figure 13**.

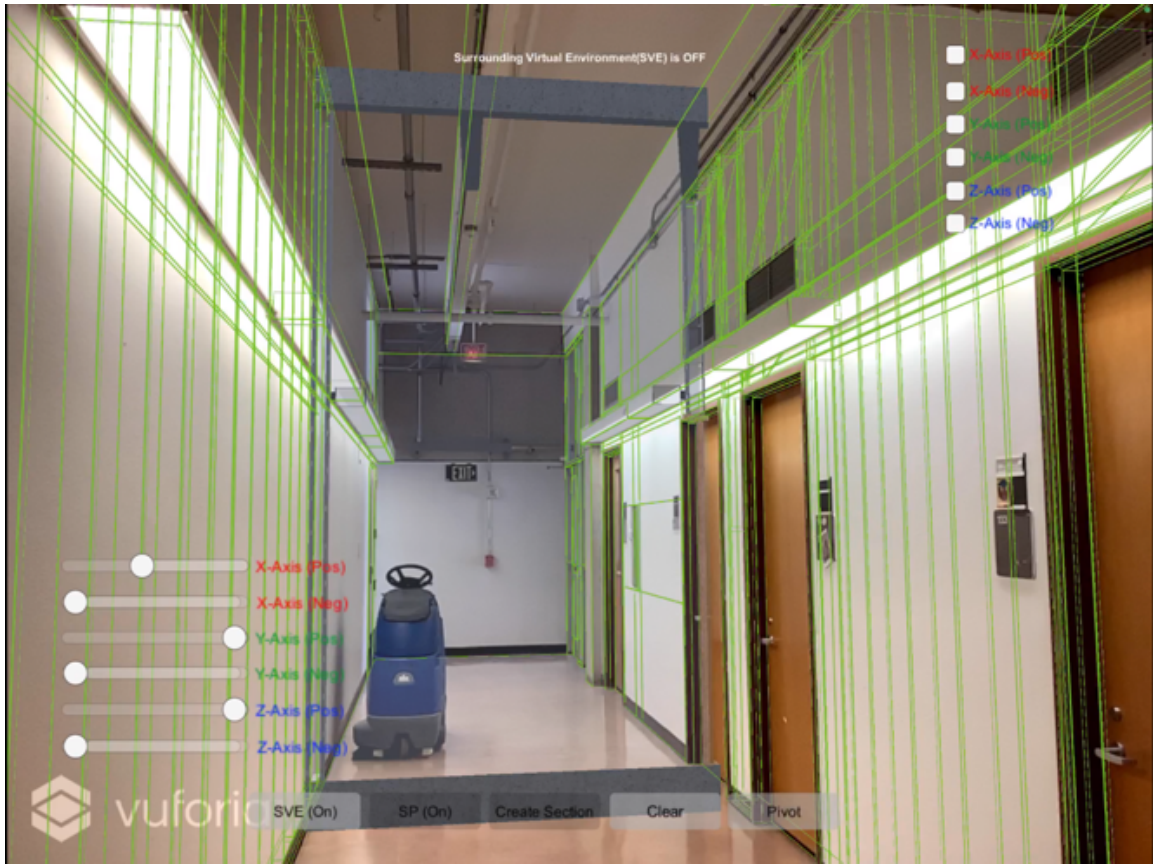

**Figure 13:** A section view along the X-Axis (parallel to the left and right walls). Sliders on the bottom left corner control the location and the orientation of the sectional plane.

We also wanted to support the touch feature and metadata retrieval during the section creation mode, through the section poche. Moreover, this feature becomes valuable when a building component consists of multiple elements, e.g., a wall with multiple layers between the two wall surfaces. Since the virtual model is not modified in terms of geometry when a section is created (a section poche is added onto the wireframe virtual model), highlighting a building component or one of its elements becomes problematic because of how Unity handles ray casting. For example, if a user wants to touch a building object through its poche, the casted ray will hit the first object it will collide with and return its ID or name. Depending on the location of the user in the environment, the ray might hit first the object (rendered invisibly) in front of the poche and eventually highlights the wrong building component or element. To overcome this problem, we adopted a solution [64] which sorts all the objects that were hit after a ray is cast from the AR camera towards the objects. The solution sorts the hit objects by their distance from the AR camera and checks which hit object is located at the sectional plane and confirms its normal direction. Using the angle between the normal vectors of the hit surface (poche and building object surfaces) and the casted ray, if the angle is small, then it is the poche, otherwise, it is the building object surface.

At the section creation mode, the user interface displays six toggles, each of which represents a sectional plane and its orientation. The user must choose one of these toggles to enable the section poche touch feature according to the section view the user is working on. If a building object is highlighted from the section poche, only the part behind the sectional plane will be highlighted, and it will be rendered with a red solid shader. The poche can accommodate multiple patterns (shaders) where each one represents an element (layer), as shown in **Figure 14**. The UI design allows every single BIM component to be selected and highlighted for examination, even if the AR device screen (iPad) has a very limited area for user interaction.

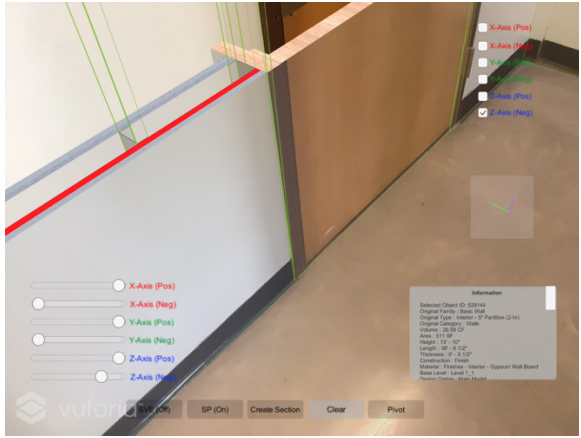

**Figure 14:** Selection of building objects or elements (layers) is enabled through the section poche. The selected element is highlighted with a red shader and its metadata are retrieved and displayed in the right bottom corner of the screen. Notice the pivot's orientation (located on the right side of the screen) is aligned with the virtual model's orientation.

### 3.1.4  Spatial and Context Awareness

Another consideration is to enable the user to understand the orientation and the coordinate system of the virtual model with respect to its location in the physical environment. To achieve this, a three-axis pivot (X, Y, and Z) has been added to the side of the UI and its orientation is frequently updated with respect to the AR camera. The pivot becomes handy when the user switches to the Create Section mode, as it allows the user to understand the location and orientation of the sectional planes **Figure 14**.

Another consideration is to render the context space behind the physical objects being sliced, so that the virtual context space (e.g., a room behind the wall) becomes visible through the "cut openings" on the physical building, while the uncut portion of the physical building component (e.g., the wall) occludes parts of the virtual context space, as shown in **Figure 15**. This effect produces a new mixed mode of real and virtual worlds that has not been exhibited in the literature before. The highly accurate registration of BIMxAR facilitates this user interface design – otherwise misaligned virtual and physical rooms/walls will not help understand the spatial relationship. In **Figure 15**, the walls, floors, and soil are rendered virtual models, instead of physical building objects. The virtual models are rendered to reveal the spaces behind the physical building as if the physical building is physically sliced (while they are not). This is an **innovative** and improved visualization compared with **Figure 13**, in which the relationship between the virtual sections and the physical building does not appear to be natural. For example in **Figure 13**, the portions of the physical door, walls and T-beam in front of the section poche is still visible, but in reality, if these physical building objects are cut to show the poche, the front portions of these physical building objects should not be visible, instead, the spaces behind them should be partially visible (as achieved in **Figure 15**). We expect that the utilization of the new mode enables a better understanding of the physical context or BIM components being explored and enhances spatial awareness.

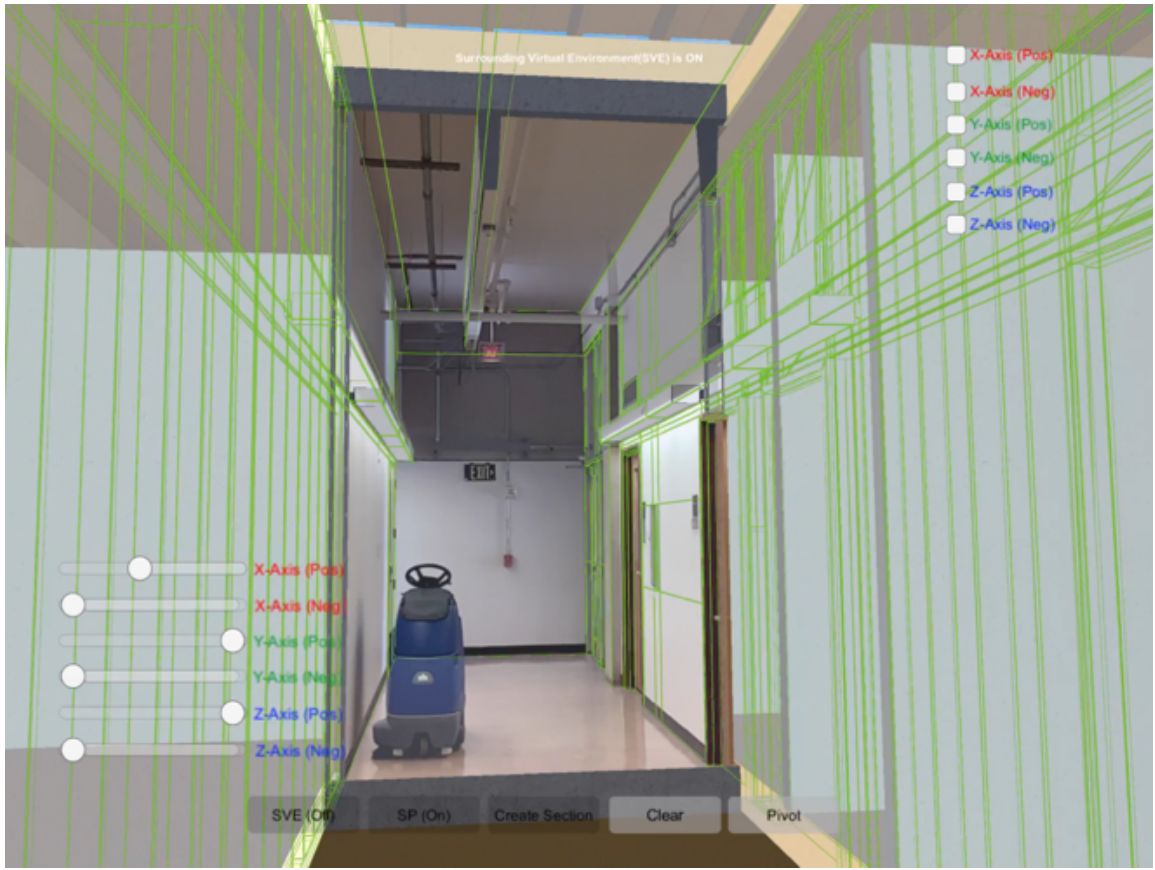

**Figure 15:** Section view revealing the spaces behind the physical objects being virtually sliced. The walls, floor, and soil in front of the section poche are rendered virtual models, instead of physical building objects.

### 3.2 Pilot User Study

The main focus of the pilot study is to measure the participants' learning gain in the mental cutting abilities and the understanding of the architectural representations, specifically section views, after using the BIMxAR prototype. Moreover, the pilot study provides us a preliminarily evaluation of the BIMxAR prototype. We used the standardized test "Santa Barbara Solids Test (SBST)" to measure the learning gain in the mental cutting ability [65]. The SBST consists of 30 questions, where each question tests the participants' mental cutting ability of single and intersecting objects.

Also, we developed a customized test "Architectural Representations Test (ART)" to measure the participants' learning gain in understanding architectural representations **Figure 16**. ART consists of 14 questions. The first half of the test measures the participant's understanding of schematic 2D section views of architectural spaces **Figure 16** (**Top**), while the other half tests the participant's understanding of detailed section views of building objects **Figure 16** (**Bottom**).

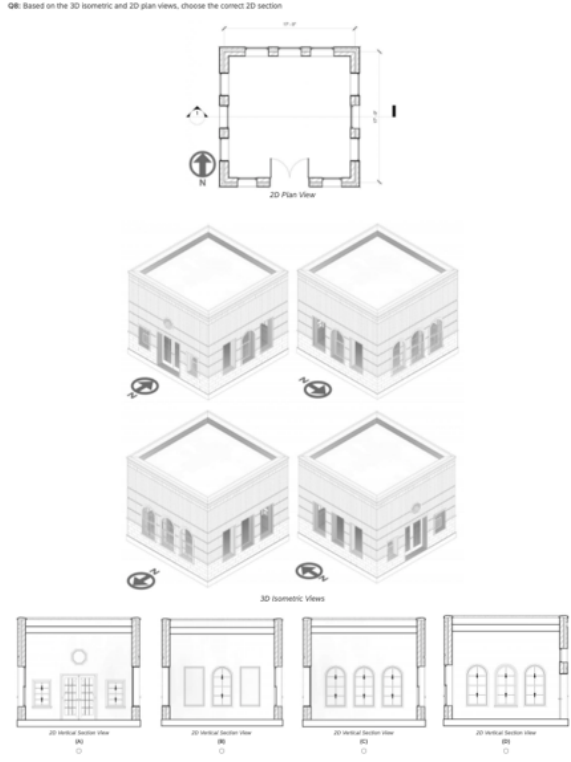

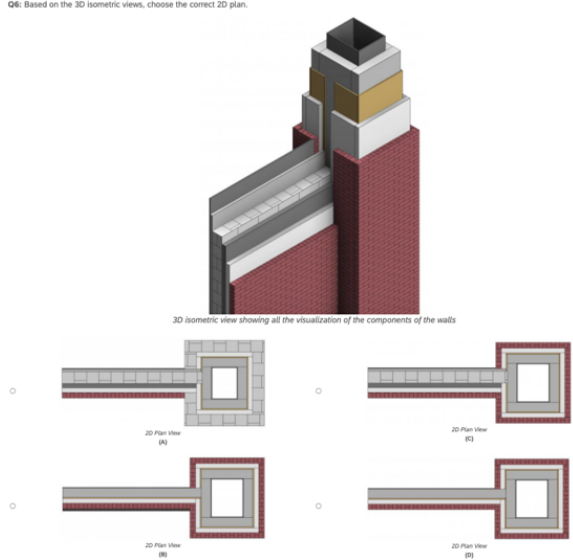

**Figure 16:** Two samples of the Architectural Representations Test (ART), developed by the authors. Top: sample question of the ART— choosing the correct schematic 2D section view of an architectural space using a 2D plan view and 3D isometric views; Bottom: sample question of the ART — choosing the correct detailed 2D section view of the architectural object(s) using a 3D isometric view that reveals all the elements of the architectural object(s).

We also measured the mental load of participants while using BIMxAR by utilizing the NASA Task Load Index (TLX) survey [66]. NASA TLX measures different demand factors of a system including temporal, physical, and mental demands, frustration, effort, and performance [66]. Additionally, the pilot sessions were video recorded for analysis and to provide more insights and explanations regarding the collected data. Prior to the pilot study, an IRB approval (IRB Number: IRB2020-1199M) has been obtained for human subject research.

### 3.2.1 Procedure

The study consisted of three phases. In Phase-I (Pretest), participants reviewed and signed the informed consent form to participate in the study, then followed by a demographical data survey. The survey collected the participants' information including their age, gender, major, degree, program, year, level of experience in (1) building construction systems and materials, (2) Building Information Modeling (BIM), and (3) Augmented Reality (AR). **Table 2** shows the participants' demographical information. Once the demographical data were collected, the participants were tested using SBST and ART. In Phase-II (Learning), the participants watched an instructional video about the BIMxAR and how to use its user interface and its functions, followed by a learning session where participants used the BIMxAR prototype and learned how to create and observe section views. In Phase-III (Posttest), the participants were tested again using SBST and AR and then followed by the NASA TLX survey. The informed consent form, demographical data survey, SBST and ART were completed through an online platform - Qualtrics.

Table 2: Participants Demographics Table of the 8 participants including their age, gender, major, level of experience in (1) building construction systems and materials, (2) building information modeling, and (3) augmented reality

| No | Age | Gender | Major | Building Construction Systems and Materials (%) | BIM (%) | AR (%) |
|---|---|---|---|---|---|---|
| 1 | 35 | F | Land. Arch. | 85.71 | 57.14 | 28.57 |
| 2 | 32 | M | Arch. | 85.71 | 85.71 | 14.29 |
| 3 | 38 | F | Arch. | 85.71 | 42.86 | 85.71 |
| 4 | 34 | F | Arch. | 57.14 | 85.71 | 100.00 |
| 5 | 30 | M | Arch. | 85.71 | 71.43 | 0.00 |
| 6 | 32 | M | Arch. | 42.86 | 42.86 | 71.43 |
| 7 | 38 | M | Arch. | 71.43 | 71.43 | 57.14 |
| 8 | 34 | F | Arch. | 71.43 | 14.29 | 14.29 |

### 3.2.2 Participants

Originally nine participants were recruited from the Department of Architecture at Texas A&M University. Among the recruited participants, eight ($n = 8$) were able to complete the entire study. All the participants (4 males, 4 females) were graduate students in the Architecture Ph.D. Program. The age range was from 30 to 38 years old with a mean age of 34 years old. The level of experience in building construction systems and materials, BIM, and AR were measured based on a 7-point Likert scale and then scaled to 100. The mean of the level of experience in building construction systems and materials, BIM, and AR was 73%, 59%, and 46% respectively **Table 2**.

### 3.3 Analysis

Due to the small number of samples and the non-normality of the data in some instances, descriptive statistical analyses, and non-parametric statistical analyses, such as the Sign test and the Wilcoxon matched-pairs signed-rank test [67], were used. The Sign and the Wilcoxon matched-pairs signed-rank tests were performed using the statistical package "JMP".

### 3.3.1 Learning Gain

The learning gain is defined as the difference between the participant scores in the pretest and posttest. The learning session was self-learning-based following instructions in the beginning and the session time ranged from 4.72 to 20.00 minutes with a mean of 13.03 minutes. **Figure 17** demonstrates the participants' scores in the SBST and the ART during the pretest and posttest phases.

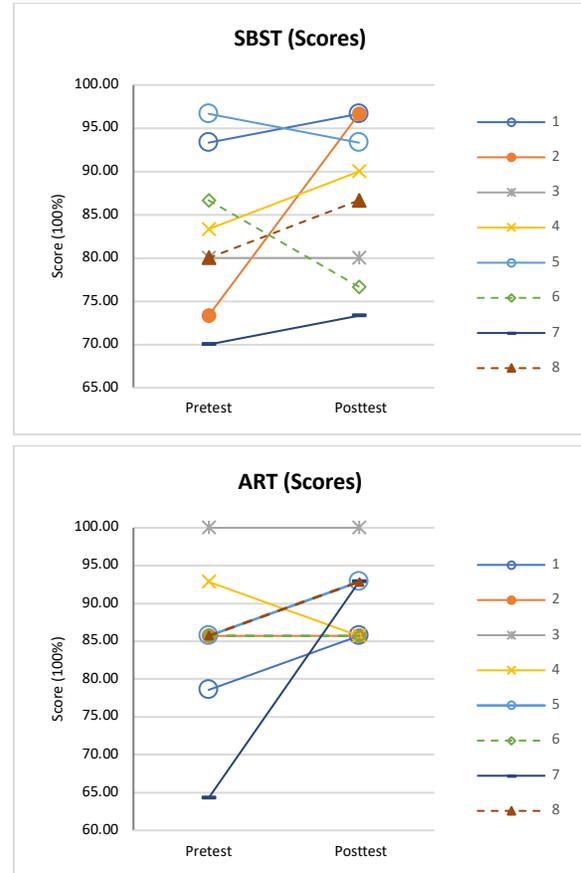

**Figure 17:** Top: participants' scores in the SBST (pretest and posttest); Bottom: participants' scores in the ART (pretest and posttest). Each participant is color-coded differently.

**Figure 17** (**Top**) shows that five (5) of the participants' scores have improved in the posttest session, and three (3) of the participants' scores did not improve. One (1) of the participants, who did not improve, had the same score in the pretest and posttest sessions. **Figure 17** (**Bottom**) indicates that four (4) of the participants' scores have improved, and four (4) of the participants' scores did not improve. Three (3) of the participants, who did not improve, had the same scores in the pretest and posttest sessions. **Table 3** shows the maximums, minimums, and means of the SBST and the ART

scores in the pretest and posttest sessions. The mean scores in the SBST (mean $_{Pretest}$ = 82.92, mean $_{Posttest}$ = 86.67) improved by 4.52%. The mean scores in the ART (mean $_{Pretest}$ = 84.82, mean $_{Posttest}$ = 90.18) improved by 6.32%. The results from the Sign test for the SBST and the ART scores with a significance level of 0.05 revealed no significant differences between the pretest and posttest sessions (p = 0.3281 and p = 0.3750 for SBST and ART respectively). Additionally, the results from the Wilcoxon matched-pairs signed-rank test with a significance level of 0.05 revealed no significant differences between the pretest and posttest sessions (p = 0.4531 and p = 0.3125 for SBST and ART respectively). However, considering both test scores and test completion times for a more comprehensive evaluation, while the SBST analysis showed an insignificant improvement, the ART analysis showed a significant improvement, as described in later subsections.

Table 3: Maximums, minimums, and means of the SBST and the ART scores in the pretest and posttest sessions.

|  | SBST | | | ART | | |
| --- | --- | --- | --- | --- | --- | --- |
|  | Pretest Score (%) | Posttest Score (%) | Score Improvement (%) | Pretest Score (%) | Posttest Score (%) | Score Improvement (%) |
| Min. | 70.00 | 73.33 | 4.76 | 64.29 | 85.71 | 33.33 |
| Max. | 96.67 | 96.67 | 0.00 | 100.00 | 100.00 | 0.00 |
| Mean | 82.92 | 86.67 | 4.52 | 84.82 | 90.18 | 6.32 |

### 3.3.2 Completion Time

The data collection method was enhanced during the pilot user study. Separate completion times of the SBST and the ART were originally not obtained during the pretest session but was obtained during the posttest session. Because test completion times could provide additional useful information for students' performance evaluation, we decided to conduct an analysis with the completion times based on obtained data, with reasonable assumptions. The collected data related to the completion time was the entire duration time of the pretest session, which included the time to fill and complete the informed consent form, demographical data survey, SBST, and ART, and the transition time between each test. Based on separate testing of the time for completing the consent form and demographical data survey, that time is approximately 5 minutes. Therefore, to calculate the completion time for the SBST and ART in the pretest session, 5 minutes were deducted from the pretest duration time, as calculated completion time for the pretest (i.e., CT $_{Pretest}$). Two of the participants were excluded from the completion time calculations since their total duration times in the pretest were treated as outliers. To calculate the completion times for each test in the pretest session, we applied the obtained ratios of the tests from the posttest session. **Figure 18** shows that 83.33% of the participants have completed the SBST and ART in a shorter period during the posttest session than in the pretest session. **Table 4** shows the maximums, minimums, and means of the completion time of the SBST and the ART in the pretest and posttest sessions. The mean completion times of the SBST (mean $_{Pretest}$ = 12.95 minutes, mean $_{Posttest}$ = 8.32 minutes) improved (reduced) by 35.74%. The mean completion times of the ART (mean $_{Pretest}$ = 20.87 minutes, mean $_{Posttest}$ = 12.91 minutes) improved (reduced) by 38.11%.

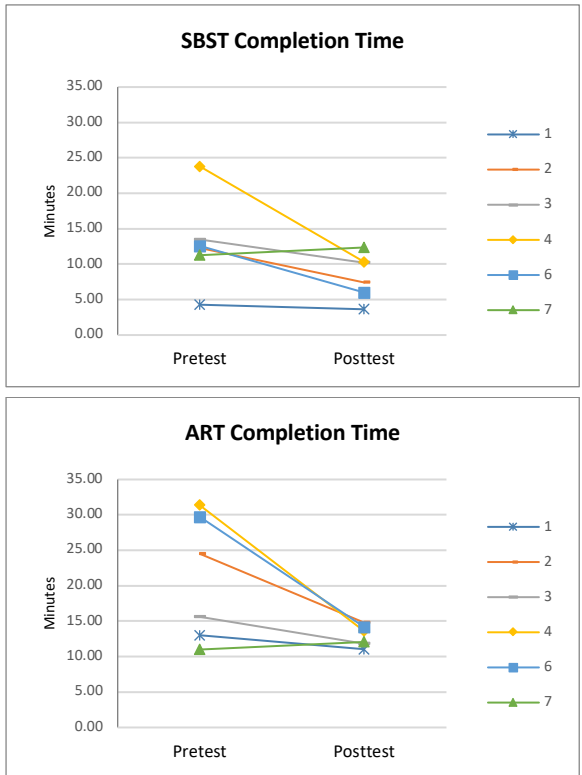

**Figure 18:** Top: Participants' completion time for SBST in the pretest and posttest sessions; Bottom: Participants' completion time for ART in the pretest and posttest sessions. Each participant is color-coded differently.

**Table 4:** Maximums, minimums, and means of the completion time (in minutes) of the SBST and the ART in the pretest and posttest sessions.

|  | SBST Completion Time | | | ART Completion Time | | |
|---|---|---|---|---|---|---|
|  | Pretest | Posttest | Completion Time Improvement (%) [Value Negated] | Pretest | Posttest | Completion Time Improvement (%) [Value Negated] |
| Min. | 4.28 | 3.63 | 15.15 | 11.00 | 11.04 | -0.41 |
| Max. | 23.80 | 12.36 | 48.06 | 31.35 | 14.79 | 52.81 |
| Mean | 12.95 | 8.32 | 35.74 | 20.87 | 12.91 | 38.11 |

### 3.3.3 Timed Scores

Computing the timed scores for the SBST and the ART during the pretest session may provide better understanding of the participants' performance and learning gains after using the BIMxAR prototype. The timed score is defined as *the test score divided by the test completion time*.

**Figure 19** (**Top**) demonstrates that all except one of the participants' SBST timed scores improved in the posttest session. On the other hand, the positive slopes in **Figure 19** (**Bottom**) show that all the participants' ART timed scores improved in the posttest session. **Table 5** shows that the mean timed score in the SBST (mean $_{pretest}$ = 8.38, mean $_{posttest}$ = 12.48) improved by 48.93%. Additionally, the mean timed score in the ART (mean $_{pretest}$ = 4.60, mean $_{posttest}$ = 7.01) improved by 52.26%.

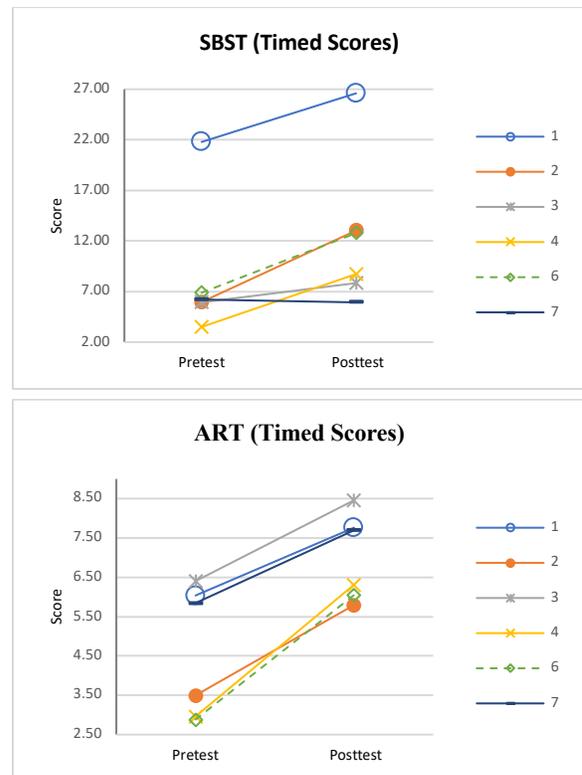

**Figure 19:** Top: participants' timed scores in the SBST (pretest and posttest); Bottom: participants' timed scores in the ART (pretest and posttest). Each participant is color-coded differently.

**Table 5:** Maximums, minimums, and means of the SBST and the ART timed scores in the pretest and posttest sessions.

|  | SBST | | | ART | | |
|---|---|---|---|---|---|---|
|  | Pretest Timed Score | Posttest Timed Score | Timed Score Improvement (%) [Value Negated] | Pretest Timed Score | Posttest Timed Score | Timed Score Improvement (%) [Value Negated] |
| Min. | 3.50 | 5.93 | 69.43 | 2.88 | 5.79 | 100.87 |
| Max. | 21.79 | 26.60 | 22.07 | 6.40 | 8.46 | 32.10 |
| Mean | 8.38 | 12.48 | 48.93 | 4.60 | 7.01 | 52.26 |

The results from the Sign test for the SBST timed scores with a significance level of 0.05 revealed no significant differences between the pretest and posttest sessions (p = 0.2188). Additionally, the results from the Wilcoxon matched-pairs signed-rank test for the SBST timed scores with a significance level of 0.05 revealed no significant difference between the pretest and posttest sessions (p = 0.0625).

On the other hand, the results from the Sign test for the ART timed scores with a significance level of 0.05 revealed a significant difference between the pretest and posttest sessions (p = 0.0313). Similarly, the results from the Wilcoxon matched-pairs signed-rank test for the ART timed scores with a significance level of 0.05 revealed **a significant improvement** from the pretest to the posttest sessions (p = 0.0313).

### 3.4 NASA TLX Survey

The NASA TLX survey was used to assess the overall workload while using the BIMxAR. NASA TLX is a multidimensional assessment survey that measures different demand factors of a system including temporal, physical, and mental demands, frustration, effort, and performance. All the demand factors, except the performance, have a positive correlation to the overall workload. The survey consists of two parts: ratings and weights. The rating section rates each demand factor independently, where the maximum possible rating is 100. The weights section has 15 pair-wise questions to compare the six demand factors, in which each question asks the user to pick the demand factor that contributes more in the paired factors to the workload. The maximum times a demand factor can be chosen is five; hence, the maximum possible weight for any demand factor is five. The adjusted rating for each demand factor is computed by, first, multiplying its demand factor rate by its corresponding demand factor weight, and then dividing by 15, thus the maximum possible adjusted rating is 33.3 (**Eq.1**). The overall workload is the sum of all adjusted ratings, where the maximum possible overall workload is 100 (**Eq.2**) [66].

$$AdjustedRating = \frac{Demand\ Factor\ Rate \times Demand\ Factor\ Weight}{15} \quad \textbf{Eq.1}$$

$$Overall\ Workload = Sum\ of\ all\ Adjusted\ Ratings \quad \textbf{Eq.2}$$

**Figure 20** shows the adjusted ratings for each demand factor per participant. Additionally, **Table 6** demonstrates the maximums, minimums, and means of all the demand factors adjusted ratings (out of 33.3). The results show that the mental demand was the highest adjusted rating (mean = 11.92), yet it is still considered to be low (35.79%). Moreover, the negated performance factor was the third-lowest adjusted rating (mean = 3.83), which means high performance perceived by the participants. Furthermore, **Figure 21** demonstrates the overall workload of each participant (out of 100). Generally, the overall workload was rated as low among all the participants (max. = 52% and mean = 34.75%).

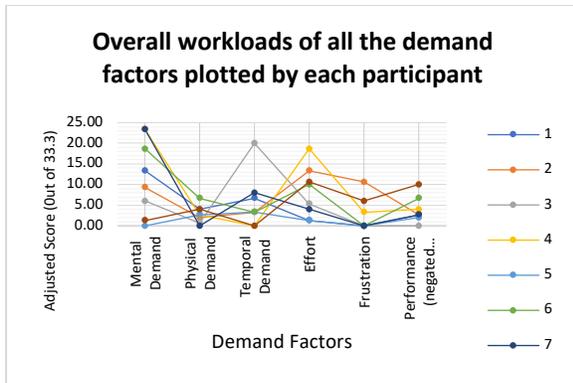

**Figure 20:** Adjusted ratings for the demand factors. Each participant is color coded differently.

**Table 6:** Maximums, minimums, and means of all demand factors adjusted ratings (out of 33.3)

|  | Mental Demand | Physical Demand | Temporal Demand | Effort | Frustration | Performance [Value Negated] |
|---|---|---|---|---|---|---|
| Max. | 23.33 | 6.67 | 20.00 | 18.67 | 10.67 | 10.00 |
| Min. | 0.00 | 0.00 | 0.00 | 1.33 | 0.00 | 0.00 |
| Median | 11.33 | 2.67 | 3.33 | 7.67 | 0.00 | 2.67 |
| Mean | 11.92 | 2.83 | 5.58 | 8.08 | 2.50 | 3.83 |

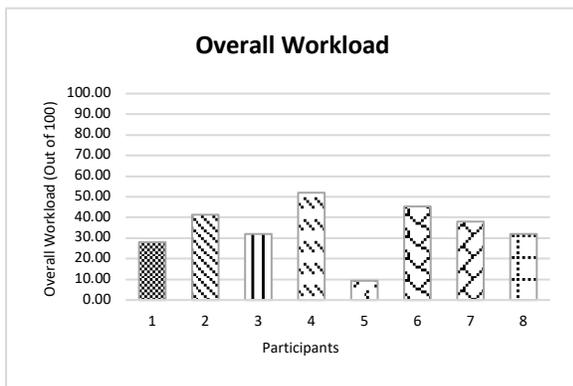

**Figure 21:** Overall workloads of all the demand factors for each participant.

### 4. Discussions

The current research showcased our working prototype (BIMxAR) towards this approach. We discussed the related technical aspects and performances of BIMxAR. We presented our workflow, a complete solution for utilizing a BIM project and its metadata in AR. We explored different registration methods in indoor environments. BIMxAR utilizes DL-3S-BIM as our registration method for the scale of buildings. Based on our experiments and analysis, the proposed method has been proven to provide the best solution in terms of accuracy and robustness with only minor errors throughout the virtual model. Based on our literature review, such a method was not found for AR registration in buildings, thus an innovation which can be utilized in AR applications in AEC.

The study has innovated AR visualization by registering architectural representations of building components in built environments and allowed users to interact with them and retrieve their BIM metadata. We explored the mechanisms for highlighting building objects with an AR interface. Also, we presented our approach to enable the user to understand the orientation and the coordinate system of the virtual model with respect to its location in the physical environment through body movement in the environment, facilitating embodied learning.

We presented an innovative method to create sections within the AR environment. The method enables the user to further inspect the building from different views through controlling the location and orientation of the sectional planes. Additionally, it allows the user to retrieve the building component's related information through the section poche. The study also developed and showcased a novel mixed-mode of real and virtual worlds (mixed reality) by revealing the spaces behind the physical objects being sliced in a section view for a better understanding of the spatial relationships in a building **Figure 15**. The highly accurate registration of BIMxAR using our registration approach, DL-3S-BIM, facilitated this mixed-mode.

In addition to the AR technology innovations, the project aimed for learning innovations in architectural education. The study presented the results of the pilot user study that was conducted to measure the participants' learning gain in subjects, including the mental cutting abilities and the understanding of the architectural representations. Eight (*n* = 8) graduate students were recruited from the Architecture Department to participate in the study. The pilot user study utilized a pretest-training-posttest design, in which the learning gain in the mental cutting ability was measured using the SBST, while the learning gain in understanding architectural representations was measured using the ART. Also, the study presented the results of the participant's mental cognitive load while using BIMxAR using the NASA TLX questionnaire.

Even though little training could have a positive impact on the participant's spatial ability [25], in the reviewed literature, participants underwent multiple lengthy sessions of training, as seen in [17], [26]. The results from the pilot user study were promising, considering the small number of samples and the short training period. We detected score improvements in the posttest sessions in the SBST and ART, although not statistically significant. However, for a more comprehensive learning evaluation, the study found clear evidence of the AR contribution in reducing the tasks' completion time. BIMxAR reduced the test completion time in the posttest session. When incorporating completion time as a factor of performance, minor improvement was detected in the SBST timed scores during the posttest session, yet not statistically significant. However, **the ART timed scores were significantly improved** during the posttest session, which is our main focus in architectural education. The difference between the improvement in the SBST and the ART timed scores during the posttest session is expected because that studying of building section views by the participants using BIMxAR is more related to ART than SBST.

By seeing the superimposed virtual information aligned with the physical world, students' extraneous cognitive load can be reduced, and ultimately the learning process would be enhanced. It was observed that the AR registration is a major feature that students tried to utilize in the learning process. The results of the NASA TLX show that the mental cognitive demand was low when using BIMxAR. We can interpret that BIMxAR may be considered an easy and convenient learning tool.

BIMxAR as an educational tool with the integrated embodied learning capabilities and advanced visualization features has never been exhibited in the literature before. BIMxAR has the potential to improve the students' spatial abilities, particularly in understanding architectural buildings and creating complex section views. Such an AR-based learning method could be utilized to benefit the education and industry in architecture, engineering, construction, maintenance, and renovations sites.

As for future work, a test case will be conducted to measure a more detailed performance of BIMxAR and the effects of our approach on the student's knowledge gain using a larger sample size. A test case has been designed and it will compare an AR experimental group with a non-AR control group in learning buildings and BIM. The non-AR control group will utilize another version of the BIMxAR, with similar visualization functions, but without the AR registration feature, as seen in **Figure 22**.

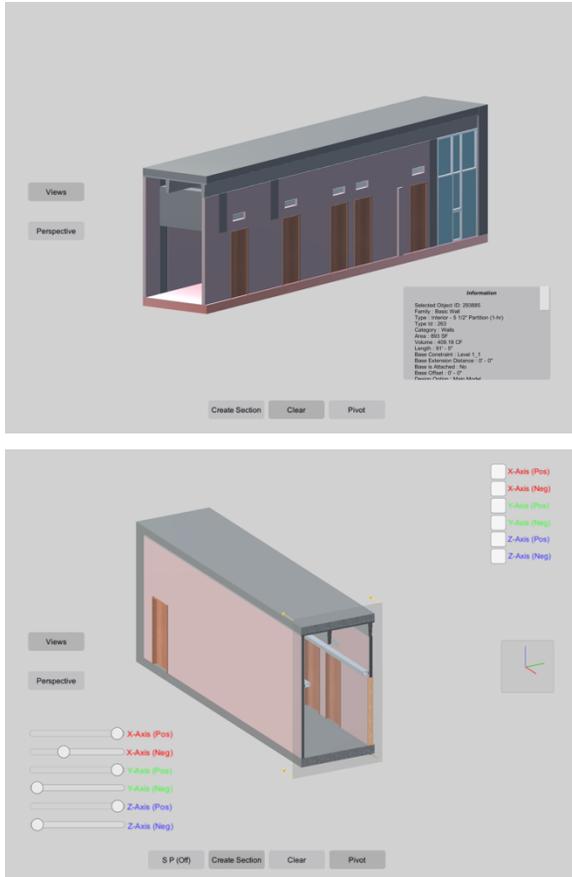

**Figure 22:** non-AR version of the BIMxAR. Left: highlighting a building object and displaying its metadata. Right: creating a section view in the section mode.

## 5. Conclusions and Future Work

The purpose of the study is to investigate new methods to improve spatial abilities in the domain of architecture education. Specifically, this research seeks to explore the AR effects on assisting students to comprehend and reproduce architectural section views. We presented our AR system prototype (BIMxAR), its highly accurate registration method (DL-3S-BIM), and its novel visualization features that facilitate the comprehension of building construction systems, materials configuration, and 3D section views of complex structures through the integration of AR, BIM, and the physical building. The study developed and showcased multiple novel AR technologies, visualization functions, and applications, as major contributions of this research:

(1) a highly accurate building-scale AR registration method (DL-3S-BIM) integrating 3D model-based Deep Learning (e.g., Vuforia Model Target), 3D-Scanning (e.g., Matterport Structured-Light), and BIM.

(2) BIM metadata retrieval in AR.

(3) virtual building section views created and registered with a physical building in AR through a full control of sectional planes' location and orientation.

(4) a mixed-mode of real and virtual worlds to show the correct spatial relationship among rooms or BIM components related to the section views.

(5) students learning building constructions and BIM with a focus on section views using AR, for which a pilot user study found promising results on the AR's potentials to improve students' spatial ability and understanding of the architectural representations.

With regard to future work, more user studies are required to draw more solid conclusions for AR's impacts on learning architectural representations. Additional test cases using other buildings that have more integrations of complex building construction systems will be conducted. Consequently, the ART's questions database will be expanded to accommodate different difficulty levels. We plan to investigate AR effects on learning energy analysis, such as daylighting analysis and computational fluid dynamics (CFD) simulations. Moreover, we will explore other AR display systems, specifically, hands-free devices, such as HoloLens, to enhance the users experience and enable additional types of interactions, e.g., eye-gaze and hand gestures. Enabled by AR registration and tracking, more user data during the learning sessions can be collected and utilized to improve learning assessments and analytics.

**ACKNOWLEDGEMENTS**

The authors thank the Office of Mapping and Space Information, Texas A&M University and Perkins and Will for the BIM (Revit) model of the building Memorial Student Center (MSC) used in the research. Also, the authors thank Dr. Mark Clayton's class and students at Texas A&M University for the BIM (Revit) model of the Langford A building used in the research. This material is based upon work supported partially by the National Science Foundation under Grant No. 2119549 and the Mattia Flabiano III AIA/Page Southerland Page Design Professorship in College of Architecture, Texas A&M University.